\documentstyle[preprint,eqsecnum,aps]{revtex}
\tightenlines
\def\sqr#1#2{{\vcenter{\hrule height.#2pt\hbox{\vrule width.#2pt height#1pt \kern#1pt \vrule width.#2pt}\hrule height.#2pt}}}

\newcommand{\loesomega}{\left[\int^\zeta g(\zeta^\prime,\sigma)d\zeta^\prime + \int^{\bar \zeta} {\bar g}({\bar \zeta}^\prime,\sigma)d{\bar \zeta}^\prime \right] \vartheta^{\hat{2}}} 

\begin{document}
\draft
\title{Plane--fronted waves in metric--affine gravity}
\author{Alberto Garc\'{\i}a$^a$\thanks{%
E-mail: aagarcia@fis.cinvestav.mx}, Alfredo Mac\'{\i}as$^{a\,b}$
\thanks{E-mail: macias@fis.cinvestav.mx, amac@xanum.uam.mx}, Dirk Puetzfeld$^c$%
\thanks{%
E-mail: dp@thp.uni-koeln.de}, and Jos\'e Socorro$^d$\thanks{%
E-mail: socorro@ifug4.ugto.mx}}
\address{$^a$ Departamento de F\'{\i}sica, CINVESTAV--IPN,\\
Apartado Postal 14--740, C.P. 07000, M\'exico, D.F., Mexico\\
$^b$ Departamento de F\'{\i}sica, Universidad Aut\'onoma\\
Metropolitana--Iztapalapa,\\
Apartado Postal 55-534, C.P. 09340, M\'exico, D.F., Mexico.\\
$^c$ Institute for Theoretical Physics, University of Cologne, \\
D--50923 K\"oln, Germany\\
$^d$ Instituto de F\'{\i}sica de la Universidad de Guanajuato,\\
Apartado Postal E-143, C.P. 37150, Le\'on, Guanajuato, Mexico.}
\date{\today}
\maketitle
%\newpage
\begin{abstract}

We study plane--fronted electrovacuum waves in metric--affine gravity
theories (MAG) with cosmological constant. Their field strengths are,
on the gravitational side, curvature $R_{\alpha}{}^{\beta}$,
nonmetricity $Q_{\alpha\beta}$, torsion $T^{\alpha}$ and, on the
matter side, the electromagnetic field strength $F$. Our starting
point is the work by Ozsv\'ath, Robinson, and R\'ozga on type N
gravitational fields in general relativity as coupled to null
electromagnetic fields. {\em File waves18.tex 2000-05-11}.
\end{abstract}


\widetext

\vspace{0.5cm}

\narrowtext
%**********************************************************

\section{Introduction}

Even though Einstein's general relativity appears almost fully corroborated
experimentally, there are several reasons to believe that the validity of
such a description is limited to macroscopic structures and to the present
cosmological era. The only available finite perturbative treatment of
quantum gravity, namely the theory of the quantum superstring \cite{quantum}%
, suggests that {\em non--Riemannian} features are present on the scale of
the Planck length. On the other hand, recent advances in the study of the
early universe, as represented by the inflationary model, involve, in
addition to the metric tensor, at the very least a {\em scalar dilaton} \cite
{inflation} induced by a Weyl geometry, i.e.\ again an essential departure
from Riemannian metricity \cite{nehe}. Even at the classical cosmological
level, a dilatonic field has recently been used to describe the presence of
dark matter in the universe as well as to explain certain cosmological
observations which contradict the fundaments of the standard cosmological
model \cite{qss}.

There is good experimental evidence that, at the present state of the
universe, the geometrical structure of spacetime corresponds to a
metric--compatible geometry in which nonmetricity, but not necessarily
the torsion, vanishes. Consequently, the full metric--affine geometry
is irrelevant for the geometrical description of the universe
today. However, during the early universe, when the energies of the
cosmic matter were much higher than today, we expect scale invariance
to prevail and, according to MAG, the canonical dilation (or scale)
current of matter, i.e. the trace of the hypermomentum current
$\Delta^\gamma{}_\gamma$ becomes coupled to the Weyl covector
$Q:=\frac{1}{4} g^{\alpha \beta} Q_{\alpha \beta} $, here
$Q_{\alpha\beta}:=-Dg_{\alpha\beta}$ is the nonmetricity of
spacetime. Moreover, shear type excitations of the material
multispinors (Regge trajectory type of constructs) are expected to
arise, thereby liberating the (metric--compatible) Riemann--Cartan
spacetime from its constraint of vanishing nonmetricity
$Q_{\alpha\beta}=0$. It is therefore important to derive and
investigate exact solutions of these theories which contain
information about the new geometric objects like torsion and
nonmetricity (for a survey of these theories see \cite{PR}).

For restricted irreducible pieces of torsion and nonmetricity, there
are similarities between the Einstein--Maxwell system and the vacuum
MAG field equations \cite{tw,oveh97}. This observation encourages us
to find new solutions for MAG theories \cite{hema99}. However, the
coupling of the post--Riemannian structures of a metric--affine
spacetime to matter is still under investigation.

The search for plane--fronted wave solutions in MAG was first
restricted to its Einstein--Cartan sector \cite
{Adam,Sippel,Hecht,Lemke}. Later, plane wave solutions with
non--vanishing nonmetricity were found by Tucker et al. \cite{TW}.
Colliding waves with the appropriate metric and an excited
post--Riemannian triplet are studied in \cite{glmms}, the
corresponding generalization to the electrovac case can be found in
\cite{gms}.

In this paper we study plane--fronted gravitational and
electromagnetic waves in metric--affine gravity theories with nonzero
cosmological constant in their {\em triplet ansatz} sector. The
plane--fronted electrovacuum--MAG waves comprize curvature,
nonmetricity, torsion, and an electromagnetic field.

The plan of the paper is as follows: In Sec.\ II, we review the
plane--fronted gravitational and electromagnetic waves in
Einstein--Maxwell theory. In Sec.\ III, we present the plane--fronted
gravitational and electromagnetic waves in MAG. In Sec.\ IV, we
specialize to particular wave solutions and, in Sec.\ V, we discuss
the results.

%****************************************************

\section{Plane--fronted gravitational and electromagnetic waves in
Einstein--Maxwell theory}

In this section we summarize the main results of Ref. \cite{orr}:
Using the null tetrad formalism, in a coordinate system
$(\rho,\sigma,\zeta, {\bar \zeta})$ (the bar denotes complex
conjugation), the metric reads \begin{equation} ds^2 = 2\left(
\vartheta^{\hat{0}} \otimes \vartheta^{\hat{1}} + \vartheta^{\hat{2}}
\otimes \vartheta^{\hat{3}}\right) \, , \label{metric} \end{equation}
with the coframe \begin{equation} \vartheta^{\hat{0}} = \frac{1}{p}\,
d \zeta \, , \quad \vartheta^{\hat{1}} = \frac{1}{p}\, d {\bar \zeta}
\, , \quad \vartheta^{\hat{2}}= - d\sigma\, , \quad
\vartheta^{\hat{3}} = \left(\frac{q}{p}\right)^2 \left( s \, d\sigma +
d\rho\right)\, , \label{coframe_rel} \end{equation} where the
structural functions $p$, $q$, and $s$ are given as follows:
\begin{eqnarray}
p(\zeta,\bar{\zeta})&=&1+\frac{\lambda_{\hbox{\scriptsize cosm}}}{6}\,
\zeta {\bar \zeta} \, , \label{p} \\ q(\sigma,\zeta,\bar{\zeta})&=&
\left(1- \frac{\lambda_{\hbox{\scriptsize cosm}}}{6}\,\zeta {\bar
\zeta}\right) \alpha(\sigma) + \zeta {\bar \beta}(\sigma) + {\bar
\zeta}\beta(\sigma) \, , \label{q} \\
s(\rho,\sigma,\zeta,\bar{\zeta})&=&-\frac{\lambda_{\hbox{\scriptsize
cosm}}}{6}\, \rho^2\alpha^2(\sigma)-\rho^2\beta(\sigma) {\bar
\beta}(\sigma) +\rho\, \partial_\sigma \left( \ln \mid q \mid \right)
+ \frac{p}{2q}\, H\left(\sigma, \zeta, {\bar \zeta}\right) \,.
\label{l}\end{eqnarray} Here $\alpha$, $\beta$, and $H$
are arbitrary functions.

Let $\tilde{R}_{\alpha \beta}$ denote the Riemannian part of the
curvature 2-form. Then we can subtract out the irreducible scalar
curvature piece
\begin{equation}
^{(6)}\tilde{R}_{\alpha \beta }:=-\frac{1}{12}(e_{\nu} \rfloor
e_{\mu} \rfloor \tilde{R}^{\nu \mu})\, \vartheta_{\alpha} \wedge
\vartheta_{\beta}  \,,
\end{equation}see \cite{compHehl}, and can define the 2-form 
\begin{equation}
S_{\alpha \beta }:=\tilde{R}_{\alpha \beta }-\,^{(6)}\tilde{R}_{\alpha \beta
  }=\,^{(1)}\tilde{R}_{\alpha \beta }+\,^{(4)}\tilde{R}_{\alpha \beta
  }={C}_{\alpha \beta }+\,^{(4)}\tilde{R}_{\alpha \beta
  }\,.
\end{equation} 
Here $e_\alpha$ denotes the (vector) frame dual to the coframe
$\vartheta^\alpha$. If the Einstein vacuum field equations (with or
without cosmological constant) are fulfilled --- in this specific case
$^{(4)}\tilde{R}_{\alpha \beta }=0$ ---, then $S_{\alpha\beta}$
becomes the Weyl conformal curvature 2-form
$C_{\alpha\beta}:=\,^{(1)}\tilde{R}_{\alpha \beta }$. Moreover, we
will introduce the propagation 1-form $k:=k_{\mu}\,\vartheta^{\mu}$
which inherits the properties of the geodesic, shear--free,
expansion--free and twistless null vector--field $k^{\mu}$
representing the propagation vector of a plane--fronted wave.

The gravitational and null electromagnetic fields are subject to the
radiation conditions
\begin{equation}
S_{\alpha\beta} \wedge k = 0\,, \qquad (e^{\alpha}\rfloor k)
\,\, S_{\alpha\beta}=0\,, \label{lob} 
\end{equation}
and
\begin{equation}
F \wedge k =0\,,\qquad (e^{\alpha} \rfloor k)\, e_{\alpha} \rfloor F =
0\,. \label{culito}
%F_{[\alpha\beta}k_{\gamma]}=0\,,\quad {\rm and}\quad k^{\alpha}F_{\alpha\beta}=0\,.  \label{culito}
\end{equation}

In the following we will solve the Einstein--Maxwell equations (for
the notion compare \cite{hema99})
\begin{eqnarray}
\frac{1}{2}\,\eta _{\alpha \beta \gamma }\wedge \tilde{R}^{\beta
\gamma }+ \lambda _{ {\rm {cosm}}}\,\eta _{\alpha }&=& \kappa \Sigma
_{\alpha }^{\rm{Max}}\,,\nonumber\\ dF&=&0\,,\nonumber\\ d{}^\ast
F&=&0\,,
\end{eqnarray}
where $\Sigma _{\alpha }^{\rm{Max}}$ represents the energy-momentum 3-form
of the Maxwell field given by 
\begin{equation}
\Sigma _{\alpha }^{\rm{Max}}:=\frac{1}{2}\left[( e_{\alpha }\rfloor F) 
\wedge {}^{\ast
}F-\left( e_{\alpha }\rfloor \,^{\ast }F\right) \wedge \,F\right].
\label{electro1}
\end{equation}
Writing the electromagnetic field as 
\begin{eqnarray}
F &=&\frac{1}{2}F_{ab}\,dx^{a}\wedge dx^{b} \nonumber \\ &=&f(\zeta
,\sigma )\,d\zeta \wedge d\sigma +\bar{f}(\bar{\zeta},\sigma )\,d
\bar{\zeta}\wedge d\sigma \,, \label{faradansatz}
\end{eqnarray}
with $f(\zeta ,\sigma )$ an arbitrary function of its arguments, one finds for
the energy-momentum 3-form of the electromagnetic field as nonvanishing 
component
\begin{equation}
\Sigma _{\hat{2} }^{\rm{Max}}= - 2\,p^{2}f{\bar{f}\;}
\vartheta ^{\hat{0}}\wedge \vartheta^{\hat{1}}\wedge \vartheta ^{\hat{2}}\,,
\end{equation}
in agreement with the result for $T_{ab}$ mentioned in
Ref. \cite{orr} Eq. (3.7).

The surfaces of constant $\sigma $ are the wave fronts of the
electromagnetic waves. The above conditions
(\ref{lob})--(\ref{culito}) restrict the function $\alpha(\sigma) $ to
the real domain whereas $\beta (\sigma )$ can be complex valued.

The function $H$, for a combined gravitational and electromagnetic
wave, has to fulfill the equation
\begin{equation}
H_{,\zeta{\bar \zeta}} + \frac{\lambda_{\hbox{\scriptsize cosm}}}{3p^2} \,
H =\frac{2 \kappa p}{q}  f {\bar f}   \, .  \label{fleq}
\end{equation}
In order to solve this non--homogeneous equation, one observes that a
complex combination of an arbitrary holomorphic function
$\Phi=\Phi(\zeta,\sigma)$ of the form
$\Phi_{,\zeta}-(\lambda_{\hbox{\scriptsize{cosm}}}/3) ({\bar \zeta}/p)\Phi$
is the general complex solution to the corresponding homogeneous
equation of (\ref{fleq}). Thus, the real $H_{{\rm h}}$--solution to
the homogeneous equation is given by
\begin{equation}
H_{{\rm h}}= \Phi_{,\zeta} - \frac{\lambda_{\hbox{\scriptsize cosm}}}{3}\, 
\frac{\bar \zeta}{p}\, \Phi + {\bar \Phi}_{,\bar \zeta} - \frac{\lambda_{%
\hbox{\scriptsize cosm}}}{3}\, \frac{\zeta}{p} \,{\bar \Phi}\, .
\end{equation}

This structure sheds light on how to find the general solution of the
non--homogeneous equation. Let us look for the particular solution
$H_{{\rm p}}$ of the form
\begin{equation}
H_{{\rm p}} = \mu_{,\zeta} - \frac{\lambda_{\hbox{\scriptsize
cosm}}}{3}\, \frac{ \bar \zeta}{p}\, \mu + {\bar \mu}_{,\bar \zeta}
-\frac{\lambda_{\hbox{\scriptsize cosm}}}{3}\, \frac{\zeta}{p}\, {\bar
\mu} \, , \label{machete}
\end{equation}
where $\mu=\mu(\sigma,\zeta, {\bar \zeta})$, such that the function 
\begin{equation}
H_{(1)} := \mu_{,\zeta} - \frac{\lambda_{\hbox{\scriptsize cosm}}}{3}\,
\frac{\bar \zeta}{p}\, \mu \, ,
\end{equation}
satisfies the equation 
\begin{equation}
H_{(1),\zeta{\bar \zeta}} + \frac{\lambda_{\hbox{\scriptsize cosm}}}{3} \frac{%
H_{(1)}}{p^2} =\frac{\kappa p}{q} f {\bar f}\,  \, .  \label{fleq1}
\end{equation}
Then it follows that $\mu$ itself is subject to 
\begin{equation}
\left(\mu_{,\bar \zeta}\right)_{,\zeta \zeta} - \frac{\lambda_{%
\hbox{\scriptsize cosm}}}{3} \left(\frac{\bar \zeta}{p}\mu_{,\bar \zeta%
}\right)_{,\zeta} = \frac{\kappa p}{q} f {\bar f}\, \, ,  \label{fleqmu}
\end{equation}
with the general solution 
\begin{equation}
\mu = \kappa \int^{\bar \zeta} d{\bar \zeta} p^2 \int^\zeta \frac{%
d\zeta^\prime}{p^2} \int^{\zeta^\prime} d \zeta^{\prime\prime} \frac{p}{q} f 
{\bar f}\, .
\end{equation}
For any given function $f$ one integrates for $\mu $ and, by using
(\ref{machete}), one obtains $H_{{\rm p}}$. The general $H$ is
constructed simply by adding the homogeneous solution $H_{{\rm h}}$
to $H_{{\rm p}}$,
\begin{equation}
H=H_{{\rm h}}+H_{{\rm p}}\,.
\end{equation}

The general solution $H$ is characterized by the selfdual part of the
conformal Weyl 2--form
\begin{equation}
{}^{+}C_{\alpha \beta}:=\frac{1}{2}(C_{\alpha \beta}+i {}^{\ast}
C_{\alpha \beta})\,,
\end{equation}
the trace--free Ricci 1--form
\begin{equation}
\tilde{R}\!\!\!\!\!\!\nearrow_{\alpha}:=e_{\beta}\rfloor
\tilde{R}_{\alpha}{}^{\beta}-\frac{1}{4}\tilde{R}\vartheta_{\alpha}\,,
\end{equation}
the Ricci scalar
\begin{equation}
\tilde{R}:=e_{\alpha} \rfloor e_{\beta} \rfloor \tilde{R}^{\alpha \beta}\,,
\end{equation}
and the electromagnetic 2--form $F$. The ansatz
(\ref{metric})--(\ref{l}) yields
\begin{eqnarray}
^{+}C_{\hat{2}\hat{0}}
&=&-\,^{+}C_{\hat{0}\hat{2}}\,=\frac{1-i}{4}\,p\,q\left(
H_{,\zeta\zeta}+\frac{\lambda_{\rm{cosm}}}{3}\,\frac{\bar{\zeta}}{p}\,
H_{,\zeta} \right)\vartheta ^{\hat{0}}\wedge \vartheta ^{\hat{2}}\,,\\
^{+}C_{\hat{2}\hat{1}} &=&-\,^{+}C_{\hat{1}\hat{2}} \,=
\frac{1+i}{4}\,p\,q\left(
H_{,\bar{\zeta}\bar{\zeta}}
+\frac{\lambda_{\rm{cosm}}}{3}\,\frac{{\zeta}}{p}\,
H_{,\bar{\zeta}} \right)\vartheta ^{\hat{1}}\wedge \vartheta ^{\hat{2}}\,,\\
\tilde{R}\!\!\!\!\!\!\nearrow_{\hat{2}}&=&p\,q\left(H_{,\zeta \bar{\zeta}}
+\frac{\lambda_{\rm{cosm}}}{3p^2}\,H\right)
\vartheta^{\hat{2}}=
2\kappa\,p^2f\bar{f}\, \vartheta ^{\hat{2}}\,, \\
\tilde{R}&=&4\lambda_{\rm{cosm}}\, , \\
F &=&dA=-d\left[ \left( \int^{\zeta}f(\zeta ^{\prime },\sigma )d\zeta ^{\prime
  }+\int^{\bar{\zeta}}{\bar f}({\bar \zeta} ^{\prime },\sigma )d{\bar \zeta} ^{\prime }\right) \vartheta ^{\hat{2}}\right]\,.
\end{eqnarray}
The Weyl 2--form could be written still a bit more compactly according to
\begin{equation}
p\,q\left(
H_{,\zeta\zeta}+\frac{\lambda_{\rm{cosm}}}{3}\,\frac{\bar{\zeta}}{p}\,
H_{,\zeta} \right)=\partial_\zeta\left[q^2\,\partial_\zeta\!\left(
\frac{p}{q}\,H\right)\right]\,,
\end{equation}
but the form given above is more practical if a certain function $H$ 
is explicitly given and calculations need to be done. 

It is worthwhile to mention the existence of a conformally flat
solution given by 
\begin{equation} H=\frac{1}{p}\,\left(
u+{\bar{v}}\zeta +v{\bar{\zeta}}+w\zeta {\bar{\zeta}}\right) \,,
\end{equation} 
where $u$, $v$, $w$ are arbitrary and $u$, $w$ real functions of
$\sigma $. The subbranch of the studied metric with constant curvature
arises form the above expression by setting $w=-(\lambda
_{\hbox{\scriptsize cosm}}/6)u$.

If the electromagnetic field is switched off, one arrives at the
non--twisting type N solutions of Garcia et al. \cite{gaple}.

%**************************************************************

\section{Plane--fronted gravitational and electromagnetic waves in MAG}

In this section we generalize the type N gravitational and electromagnetic
waves to the metric--affine gravity theories. We will present exact
solutions of the field equations belonging to the Lagrangian 
\begin{equation}
L=V_{{\rm MAG}}+V_{{\rm Max}}\,,  \label{Ltot}
\end{equation}
where $V_{{\rm Max}}=-(1/2)F\wedge \hspace{-0.8em}{\phantom{F}}^{\ast }F$
is the Lagrangian of the Maxwell field and $F=dA$ is the electromagnetic
field strength. The MAG Lagrangian considered here reads (a more general MAG
Lagrangian can be found in \cite{hema99}):
\begin{eqnarray}
V_{{\rm MAG}}=\frac{1}{2\kappa }\,&&\left[ -a_{0}\,R^{\alpha \beta }\wedge
\eta _{\alpha \beta }-2\lambda _{\hbox{\scriptsize cosm}}\,\eta
\right.\nonumber \\ &&+\left.T^{\alpha }\wedge
{}^{\ast }\!\left( \sum_{I=1}^{3}a_{I}\,^{(I)}T_{\alpha }\right) \right. 
\nonumber \\
&&+\left. 2\left( \sum_{I=2}^{4}c_{I}\,^{(I)}Q_{\alpha \beta }\right) \wedge
\vartheta ^{\alpha }\wedge {}^{\ast }\!\,T^{\beta }\right.  \nonumber \\
&&+\left. Q_{\alpha \beta }\wedge {}^{\ast }\!\left(
\sum_{I=1}^{4}b_{I}\,^{(I)}Q^{\alpha \beta }\right) \right. \nonumber \\
&&+\left. b_{5}(^{(3)}Q_{\alpha \gamma }\wedge {}\vartheta ^{\alpha
}\!)\wedge ^{\ast }(^{(4)}Q^{\beta \gamma }\wedge \vartheta _{\beta })\right]
\nonumber \\
&&\hspace{-0.6cm}-\frac{1}{2\rho}\,R^{\alpha \beta }\wedge {}^{\ast }\left(
z_{4}\,^{(4)}Z_{\alpha \beta }\right) \,.  \label{lobo}
\end{eqnarray}
where 
\begin{equation}
a_{0},\ldots, a_{3}, b_{1},\ldots, b_{5}, c_{2},c_{3},c_{4},z_{4}\,,
\label{consts}
\end{equation}
are dimensionless coupling constants, $\kappa $ is the weak and $\rho$
the strong gravitational coupling constant. The cosmological constant
is denoted by $\lambda _{\hbox{\scriptsize cosm}}$.
The signature of spacetime is $(-+++)$, the volume 4--form $\eta
:={}^{\ast }\!\,1$, the 2--form $\eta _{\alpha \beta }:=\,^{\ast
}(\vartheta _{\alpha }\wedge \vartheta _{\beta })$.

The two MAG field equations for electromagnetic matter are given by
\cite{PR}
\begin{eqnarray}
DH_{\alpha }-E_{\alpha } &=&\Sigma _{\alpha }^{\rm Max}\,,  \label{first} \\
DH^{\alpha }\,_{\beta }-E^{\alpha }\,_{\beta } &=&0\,,  \label{second}
\end{eqnarray}
with $\Sigma _{\alpha }^{\rm{Max}}$ as defined in (\ref{electro1}). It
can be alternatively written as
\begin{equation}
\Sigma _{\alpha }^{\rm{Max}}= e_{\alpha }\rfloor V_{{\rm Max}%
}+(e_{\alpha }\rfloor F)\wedge H \,.  \label{electro}
\end{equation}

For the torsion and nonmetricity field configurations, we concentrate on the
simplest non--trivial case {\em with} shear. According to its irreducible
decomposition \cite{PR}, the nonmetricity contains two covector pieces,
namely the dilation piece
\begin{equation}
^{(4)}Q_{\alpha\beta}= Q\,g_{\alpha\beta}
\end{equation}
and the proper shear piece
\begin{equation}
^{(3)}Q_{\alpha\beta}={\frac{4}{9}}\left(\vartheta_{(\alpha}e_{\beta)}%
\rfloor \Lambda - {\frac{1}{4}} g_{\alpha\beta}\Lambda\right)\,,\qquad %
\hbox{with}\qquad \Lambda:= \vartheta^{\alpha}e^{\beta}\rfloor\! {%
\nearrow\!\!\!\!\!\!\!Q}_{\alpha\beta} \,.  \label{3q}
\end{equation}
Accordingly, our ansatz for the nonmetricity reads
\begin{equation}
Q_{\alpha\beta}=\, ^{(3)}Q_{\alpha\beta} +\, ^{(4)}Q_{\alpha\beta}\,.
\label{QQ}
\end{equation}
The torsion, in addition to its tensor piece, encompasses a covector and an
axial covector piece. Let us choose only the covector piece as
non--vanishing: 
\begin{equation}
T^{\alpha}={}^{(2)}T^{\alpha}={\frac{1}{3}}\,\vartheta^{\alpha}\wedge T\,,
\qquad \hbox{with}\qquad T:=e_{\alpha}\rfloor T^{\alpha}\,.  \label{TT}
\end{equation}
Thus we are left with the three non--trivial 1--forms $Q$, $\Lambda$,
and $T$. We shall assume that this triplet of 1--forms shares the
spacetime symmetries, that is, its members are proportional to each
other \cite {heh96,he96,PLH97,ma,slmm98,ghlms}.  Our ansatz for the
nonmetricity is expected to require a nonvanishing post--Riemannian
term quadratic in the segmental curvature. This is the term in
(\ref{lobo}) carrying the coupling constant $z_4$ (note that the
enumeration of the constants stems from the general Lagrangian
mentioned in \cite{hema99}).

We assume the following so--called {\em triplet ansatz} for our
three 1--forms in (\ref{QQ}) and (\ref{TT}), 
\begin{equation}
Q=k_{0}\,\omega \,,\qquad \Lambda =k_{1}\,\omega \,,\qquad T=k_{2}\,\omega
\,,  \label{tripp}
\end{equation}
where $k_{0}$, $k_{1}$, and $k_{2}$ are constants. The triplet ansatz
(\ref{tripp}) reduces the {\em electrovacuum} MAG field equations
(\ref{first})--(\ref{second}) to an effective Einstein--Proca--Maxwell
system:
\begin{eqnarray}
\frac{a_{0}}{2}\,\eta _{\alpha \beta \gamma }\wedge \tilde{R}^{\beta \gamma
}+\lambda _{{\rm {cosm}}}\,\eta _{\alpha } &=& \kappa \,\left[ \Sigma
_{\alpha }^{(\omega )}+\Sigma _{\alpha }^{\rm{Max}}\right] \,\label{fieldob1}, \\
d\,{}^{\ast}d\omega + m^{2}\,{}^{\ast}\omega&=&0 \,, \\
dF =0 \quad &,&\quad d { }^{*} F=0 \,.  \label{fieldob}
\end{eqnarray}
These are partial differential equations in terms of the coframe
$\vartheta^\alpha$, the triplet 1--form $\omega $, and the
electromagnetic potential 1--form $A$; here the tilde $\tilde{%
\null}$ denotes again the Riemannian part of the curvature. The
energy--momentum current of the triplet field $\omega$ reads
\begin{eqnarray}
\Sigma _{\alpha }^{(\omega )} &:&=\frac{z_{4}k_{0}^{2}}{2\rho }\left\{
\left( e_{\alpha }\rfloor d\omega \right) \wedge {}^{\ast }d\omega -\left(
e_{\alpha }\rfloor \,^{\ast }d\omega \right) \wedge \,d\omega \right. 
\nonumber \\
&&\quad \left. +\;m^{2}\,\left[ (e_{\alpha }\rfloor \omega )\wedge {}^{\ast
}\omega \;+\;(e_{\alpha }\rfloor \,^{\ast }\omega )\wedge {}\omega \right]
\right\} \,;  \label{ProcaEM}
\end{eqnarray}
the effective ``mass'' $m$ depends, additionally, on $\kappa $ and the
strong gravitational coupling constant $z_{4}/\rho $, see \cite{oveh97}.

Therefore, as mentioned above, in the framework of the triplet ansatz,
the electrovacuum sector of MAG reduces to an effective
Einstein--Proca--Maxwell system. Moreover, by setting $m=0$, the
system acquires the following constraint among the coupling constants
$k_{0}$, $k_{1}$, $k_{2}$ of the triplet ansatz (\ref {tripp}) and the
constants of the Lagrangian (\ref{lobo}):
\begin{eqnarray}
-4 b_4 + \frac{3}{2} a_0 + \frac{k_1}{2 k_0}
(b_5-a_0)+\frac{k_2}{k_0}(c_4+a_0)=0 \,.\label{bedingungm0}
\end{eqnarray}
The coframe we will consider is of the form (\ref{coframe_rel}), i.e.,
it is the same as in the general relativistic case. Note that we
changed the name of the function $H$ in $s$ (cf.(\ref{l})) into
$\cal{H}$ in order to distinguish the general relativistic from the
MAG case.

Now ${\cal H}$, representing a combined gravitational MAG plane wave
and an electromagnetic wave, has to fulfill the equation
\begin{equation}
{\cal H}_{,\zeta{\bar \zeta}} + \frac{\lambda_{\hbox{\scriptsize
cosm}}}{3} p^{-2} {\cal H} = \frac{2 \kappa p}{q}\left[ f {\bar f} + g
{\bar g}\right] \, ,
\label{fleqmag}
\end{equation}
where $f=f(\zeta,\sigma)$ and $g=g(\zeta,\sigma)$ are arbitrary functions of
their arguments.

The general solution of this equation is given by ${\cal H}_{{\rm h}} + 
{\cal H}_{{\rm p}}$ with
\begin{equation}
{\cal H}_{{\rm h}}= \Phi_{,\zeta} - \frac{\lambda_{\hbox{\scriptsize cosm}}}{%
3} \frac{\bar \zeta}{p} \Phi + {\bar \Phi}_{,\bar \zeta} - \frac{\lambda_{%
\hbox{\scriptsize cosm}}}{3} \frac{\zeta}{p} {\bar \Phi}\, , 
\label{culon}
\end{equation}
and 
\begin{equation}
{\cal H}_{{\rm p}} = M_{,\zeta} - \frac{\lambda_{\hbox{\scriptsize cosm}}}{3}
\frac{\bar \zeta}{p} M + {\bar M}_{,\bar \zeta} -\frac{\lambda_{%
\hbox{\scriptsize cosm}}}{3} \frac{\zeta}{p} {\bar M} \, .
\label{machetemag}
\end{equation}
Here $M=M(\sigma,\zeta, {\bar \zeta})$ is a solution of the non--homogeneous
equation for ${\cal H}$, which is given by 
\begin{equation}
M= \kappa \int^{\bar \zeta} d{\bar \zeta} p^2 \int^\zeta \frac{%
d\zeta^\prime}{p^2} \int^{\zeta^\prime} d \zeta^{\prime\prime} \frac{p}{q} %
\left[f {\bar f} + g {\bar g}\right] \, .  \label{culiton}
\end{equation}
For given functions $f$ and $g$, one integrates (\ref{culiton}) for
$M$ and obtains ${\cal H}_{{\rm p}}$ from (\ref{machetemag}). The
general solution is obtained by adding the homogeneous solution
(\ref{culon}), where $\Phi$ is an arbitrary holomorphic function of
$\zeta$ and $\sigma$. The 1--form $\omega$ entering the triplet ansatz
(\ref{tripp}) is given by
\begin{equation}
\omega=-\left[\int^\zeta g(\zeta^\prime,\sigma)d\zeta^\prime + \int^{\bar %
\zeta} {\bar g}({\bar \zeta}^\prime,\sigma)d{\bar \zeta}^\prime \right]
\vartheta^{\hat{2}} \, ,  \label{tripp1}
\end{equation}
where $g=g(\zeta,\sigma)$ represents an arbitrary function of the coordinates. 
Moreover, the electromagnetic 2--form is given by 
\begin{equation}
F= dA = - d\left[\left(\int^\zeta
f(\zeta^\prime,\sigma)d\zeta^\prime + \int^{\bar \zeta}{\bar f}({\bar \zeta}%
^\prime,\sigma)d{\bar \zeta}^\prime \right)\vartheta^{\hat{2}}\right] \, 
\label{cuas}
\end{equation}
in terms of the arbitray function $f=f(\zeta,\sigma)$.
Inserting this ansatz into the field equations
(\ref{fieldob1})-(\ref{fieldob}) yields the following additional constraints among the
constants of (\ref{lobo}):
\begin{eqnarray}
a_0=1 \quad,\quad z_4=\frac{\rho}{2 k_0}\,.
\end{eqnarray}

%*************************************

\section{Particular solutions}

For better understanding, let us look for certain families of
particular solutions of our dynamical system by integrating
(\ref{fleqmag}) restricted to $\alpha = 1$ and $ \beta =0$. Now the
coframe in terms of $p(\zeta ,\bar{\zeta}),q(\zeta ,\bar{\zeta})$ and
${\cal H}(\sigma ,\zeta ,\bar{\zeta})$ reads
\begin{eqnarray}
\vartheta ^{\hat{0}} &=&\frac{1}{p}\,d\zeta \quad ,\quad \vartheta
^{\hat{1} }=\frac{1}{p}\,d\bar{\zeta}\quad ,\quad \vartheta
^{\hat{2}}=-d\sigma \quad , \nonumber \\ \vartheta ^{\hat{3}}
&=&\left( \frac{q}{p}\right) ^{2}\left[ \left( \frac{p}{ 2\,q}{\cal
H}(\sigma ,\zeta ,\bar{\zeta})-\frac{\lambda_{\rm cosm} }{6}\,\rho^{2}
\right)d\sigma +d\rho \right] .  \label{partcoframe}
\end{eqnarray}
Here $p$ and $q$ take the explicit form:
\begin{equation}
p(\zeta,\bar{\zeta})=1+\frac{\lambda_{\rm cosm}}{6}\zeta\bar{\zeta}\,
, \qquad q(\zeta,\bar{\zeta})=1-\frac{\lambda_{\rm
cosm}}{6}\zeta\bar{\zeta}\,.
\end{equation}
Eq.(\ref{fleqmag}) is a linear equation, therefore, one can look
independently for solutions of the non--homogeneous equation for the
$f$ exitations (associated with the electromagnetic field) and for the
$g$ exitations (associated with the post--Riemannian
pieces). Consequently, the addition of these solutions, corresponding
to $f$ and $g$, will be again a solution. For simplicity, we shall
restrict ourselves to the case where $g(\zeta,\sigma)$ and
$f(\zeta,\sigma)$ are polynomial functions of $\zeta$ and
$\zeta^{-1}$.  Let us try the cases
\begin{equation}
f\left(\zeta,\sigma\right) = f_0 \zeta^n\, , \quad n=0,\pm 1, \pm 2, \pm3,
\cdots \, .
\end{equation}
Then one obtains the following branches of solutions for ${\cal H}_{{\rm p}}$:
\newline
\underline{(i) $n<-1$} 
\begin{eqnarray}
{\cal H}_{\text{p}} &=&\frac{2 \kappa p f_{0}^2}{q}\left( \frac{(\zeta \bar{\zeta})^{1+n}}{(1+n)^{2}}%
+4\left( \frac{\lambda_{\rm cosm} }{6}\right) ^{-n-1}\ln \left| \,q\,\right| -4\left( 
\frac{\lambda_{\rm cosm} }{6}\right) ^{-n-1}\ln \left| \,p-1\,\right| \right.  \nonumber \\
&&\left. +4\sum_{r=1}^{-n-1}\frac{\left( \frac{\lambda_{\rm cosm} }{6}\right) ^{-n-r-1}%
}{r\,\left( \zeta \bar{\zeta}\right) ^{r}}\right) + \frac{8 \kappa f_{0}^2 \left( \zeta \bar{%
\zeta}\right) ^{n+1}}{(1+n)\,p},  \label{hpkleinerSOL}
\end{eqnarray}
\underline{(ii) $n=-1$} 
\begin{equation}
{\cal H}_{\text{p}}=\frac{2 \kappa f_{0}^2}{p}\left( 4\,q\,\ln \left| \,q\,\right| +\frac{2\lambda_{\rm cosm}
\zeta \bar{\zeta}}{3}\ln \left( \zeta \bar{\zeta}\right) +\frac{q}{2}\ln
^{2}\left( \zeta \bar{\zeta}\right) \right) ,  \label{hpgleichSOL}
\end{equation}
\newline
\underline{(iii) $n>-1$ } 
\begin{equation}
{\cal H}_{\text{p}}=\frac{8 \kappa f_{0}^2 q}{p}\left(
\frac{\lambda_{\rm cosm} }{6}\right) ^{-n-1}\left( \ln \left|
\,q\,\right| +\sum_{r=1}^{n}\frac{\left( _{r}^{n}\right) }{r}\left(
\left( p-2\right) ^{r}-(-1)^{r}\right) \right) +\frac{2 \kappa f_{0}^2
(\zeta \bar{\zeta})^{n+1}}{ p\,(n+1)^{2}}(4(n+1)+q).  \label{hpgroesserSOL}
\end{equation}
Similarily one can proceed with solutions for $g$, 
\begin{equation}
g\left(\zeta, \sigma\right) = g_0\, \zeta^l\, , \quad l=0,\pm 1, \pm 2, \pm3,
\cdots
\end{equation}
The form of the different branches of ${\cal H}_{{\rm p}}$ do not
change, but the substitution $n \rightarrow l$ and $f_0 \rightarrow
g_0$ should be performed. Therefore, one can obtain different branches
of solutions by combining the $f$ branches with the $g$ branches of
${\cal H}_{{\rm p}}$.

For these particular classes one can choose $ {\cal H}_{{\rm h}}$ as
displayed in (\ref{culon}).
Given $g(\zeta,\sigma)$ and $f(\zeta,\sigma)$ it is straightforward to
evaluate the 1--form $\omega$ of (\ref{tripp1}) and the electromagnetic
2--form of (\ref{cuas}).

This solution was checked by means of the computer algebra system
Reduce \cite{REDUCE} by applying its Excalc package \cite{EXCALC} for
treating exterior differential forms\cite{Stauffer}.

%**************************************************

\section{Discussion}

We investigated plane--fronted electrovacuum--MAG waves with
cosmological constant in the triplet ansatz sector of the
theory. These waves carry curvature, nonmetricity, torsion, and an
electromagnetic field. Apart from the cosmological constant, the
solutions contain four wave parameters, given by the functions
$\alpha(\sigma)$, $\beta(\sigma)$, $\bar{\beta}(\sigma)$ and
$\partial_\sigma {\rm ln}|q(\sigma,\zeta,\bar{\zeta})|$. Our
plane--fronted wave solutions are given in terms of three arbitrary
complex functions, i.e.\ $\Phi(\sigma,\zeta)$ associated with the
Riemannian part, $g(\sigma,\zeta)$ related to the non-Riemannian
triplet, and $f(\sigma,\zeta)$ corresponding to the Maxwell field. In
this way, we generalize the plane--fronted electrovacuum
Ozsvath--Robinson--Rozga waves.
In brief, the solution reads:
\\
\\
\begin{tabular}{|l|l|}
\hline\hline
ansatz for coframe $\vartheta ^{\hat{0}},\vartheta ^{\hat{1}},\vartheta ^{%
\hat{2}},\vartheta ^{\hat{3}}$ & (2.2) \\ \hline
arbitrary functions in coframe & $\alpha (\sigma ),\,\beta (\sigma )$ \\ 
\hline
MAG Lagrangian $V_{\mathrm{MAG}}$ and non-vanishing coupling constants & (3.2),(3.3) with (3.16),(3.23) \\ \hline
triplet ansatz for nonmetricity and torsion & $Q\sim T \sim \Lambda \sim \omega $,
cf. (3.11) \\ \hline
energy-momentum current of the Maxwell field & (2.11) resp. (2.13) \\ \hline
energy-momentum current of the triplet field & (3.15) \\ \hline
field equations & (3.12)-(3.14) \\ \hline
arbitrary function governing the vacuum solution ${\cal H}_{\rm h}$& $\Phi(\sigma,\zeta)$, cf. (3.18)
\\ \hline
arbitrary function in the electromagnetic 2-form $F$ & $f(\sigma,\zeta )$, cf. (3.22) \\ \hline
arbitrary function in the triplet 1-form $\omega $ & $g(\sigma,\zeta )$, cf. (3.21) \\ 
\hline
solution for the electromagnetic 2-form $F$ & $F$ $\,\sim\,\,-d(\int
f\,\,d\zeta +\bar{f}\,\,d\bar{\zeta})\,\,\vartheta ^{\hat{2}}$ \\ 
\hline
solution for the triplet 1-form $\omega $ & $\omega \,\,\sim\,\,-(\int
g\,d\zeta \,+\,\bar{g}\,d\bar{\zeta})\,\vartheta ^{\hat{2}}\,$ \\ 
\hline
solution for function $\cal{H}$$(\sigma ,\zeta , \bar{\zeta})$ entering
coframe & (3.18)-(3.20) \\ \hline\hline
\end{tabular}
\newline
\newline
The final form of $T^{\alpha}$ and $Q_{\alpha\beta}$ in terms of $g(\zeta^\prime,\sigma)$ reads,
\begin{eqnarray}
T^{\alpha}&=&-{\frac{k_{2}}{3}}\,  \left[\int^\zeta  g(\zeta^\prime,\sigma)d\zeta^\prime + \int^{\bar \zeta} {\bar g}({\bar \zeta}^\prime,\sigma)d{\bar \zeta}^\prime \right]
\vartheta^{\alpha}\wedge \vartheta^{\hat{2}} \, , \\
Q_{\alpha\beta}&=&-\frac{4k_{1}}{9} \vartheta_{(\alpha} e_{\beta)} \rfloor \loesomega \nonumber \\
&& +g_{\alpha \beta}\left(\frac{k_{1}}{9}-k_{0}\right) \loesomega \, .
\end{eqnarray}
The electromagnetic potential 1-form is given by
\begin{equation}
A=- \left( \int^{\zeta}f(\zeta ^{\prime },\sigma )d\zeta ^{\prime
  }+\int^{\bar{\zeta}}{\bar f}({\bar \zeta} ^{\prime },\sigma )d{\bar \zeta} ^{\prime }\right) \vartheta ^{\hat{2}}\,.
\end{equation}
It is straightforward to perform a detailed classification \cite{kr80}
of the plane--fronted waves in MAG by carrying through a similar
analysis as the one done by Sippel and Goenner \cite{Sippel}. We leave
this, however, for future work.

%***********************************************************
\section{Outlook}

The theories of modern physics generally involve a mathematical model,
defined by a certain set of differential equations, and supplemented 
by a set of rules for translating the mathematical results into meaningful
statements about the physical world. In the case of gravity theories,
because they deal with the most universal of the physical interactions,
one has an additional class of problems concerning the influence of the
gravitational field on other fields and matter. These are often studied
by working within a fixed gravitational field, usually an exact 
solution \cite{kr80}.
In this context our plane--fronted waves solutions contribute to enhance our understanding
of some of these questions in the framework of MAG theories, in particular
the ones concerned with the gravitational radiation.

Gravitational waves \cite{schutz} have traveled almost unimpeded through 
the universe since they were generated at times as early as $10^{-24}$ sec. after
the big bang. This radiation carries information that no electromagnetic radiation
can give to us because the electromagnetic radiation is scattered countless
times by the dense material surrounding the explosion, losing in the process
most of the detailed information it might carry about the explosion.       
Beyond this, we can be virtually certain that gravitational wave spectrum
has surprises for us, clues to phenomena we never suspected. Therefore, it
is not surprising, that considerable effort is nowadays being devoted
to the development of sufficiently sensitive gravitational wave antennas.
Moreover, observing them would provide important constraints on theories of
inflation and high--energy physics.

Even though Einstein's treatment of spacetime as a Riemannian manifold
appears almost fully corroborated experimentally, there are several reasons
to believe that the validity of such a description is limited to
macroscopic structures and to the present cosmological era.
The only available finite perturbative
treatment of quantum gravity, namely the theory of the quantum
superstring \cite{quantum}, suggests that non--Riemannian features 
are present on the scale of the
Planck length.  On the other hand, recent advances 
in the study of the early universe, as represented by the inflationary
model, involve, in addition to the metric tensor, at the very least a
scalar dilaton \cite{inflation} induced by a Weyl geometry, i.e., 
again an essential departure from Riemannian metricity \cite{nehe}.  
Even at the classical cosmological level, a 
dilatonic field has recently been used to describe the presence of
dark matter in the universe, as well as to explain certain 
cosmological observations which contradicted the fundaments 
of the standard cosmological model \cite{qss}. 

Inflation is an attractive scenario for the early universe because it makes the 
large scale homogeneity of the universe easy to understand. It also provides a mechanism
for producing initial density perturbations large enough to evolve into galaxies
as the universe expands. These perturbations are accompanied by perturbation of
the gravitational field that travel through the universe, redshifting in the same
way that photons do. The perturbations arise by parametric amplification of
quantum fluctuations in the gravitational wave field that existed before the
inflation began. The huge expansion associated
with inflation puts energy into
these fluctuations, converting them into real gravitational waves with classical
amplitudes. Even if inflation did not occur, the perturbations that lead to
galaxies must have arisen in some other way, and it is possible that this alternative
mechanism also produced gravitational waves.

It is worthwhile to stress \cite{PR} the fact that we do not believe that at the
present state of the universe the geometry of spacetime is described
by a metric--affine one. We rather think, and there is good
experimental evidence, that the present--day geometry is
metric--compatible, i.e., its nonmetricity vanishes. In earlier epochs
of the universe, however, when the energies of the cosmic ``fluid''
were much higher than today, we expect scale invariance to prevail ---
and the canonical dilation or scale current of matter, the trace of
the hypermomentum current $\Delta^\gamma{}_\gamma$, is coupled,
according to MAG, to the Weyl covector $Q^\gamma{}_\gamma$. By the
same token, shear type excitations of the material multispinors (Regge
trajectory type of constructs) are expected to arise, thereby
liberating the (metric-compatible) Riemann--Cartan spacetime from its
constraint of vanishing nonmetricity $Q_{\alpha\beta}=0$ . Tresguerres
\cite{Tres3} has proposed a simple cosmological model of Friedmann
type which carries a metric-affine geometry at the beginning of the
universe, the nonmetricity of which dies out exponentially in time.
That is the kind of thing we expect.

In full, exact solutions of the
type obtained may serve well as starting point for the upcoming
analysis of gravitational wave astronomy data. In this sense it
might contribute to our understanding of light and gravitational
wave propagation in early stages of the universe.
Moreover, plane wave solutions contribute to resolve some of the 
controversies about the existence of such gravitational radiation.

%**************************************************** 
\acknowledgments
We thank Friedrich W. Hehl for useful discussions and literature hints.
This research was supported by CONACyT Grants: 28339E,
and 32138E. 

%*****************************************************

\end{document}